\documentclass[10pt]{article}
\usepackage{latexsym,amssymb}
\usepackage{amsmath,amsbsy}
\usepackage[all]{xy}
\usepackage{amsopn,amstext}
\usepackage{cite}
\usepackage{amssymb}
\usepackage{rotating}
\usepackage{amsmath}
\usepackage{amsopn,amstext}
\usepackage{pdflscape}
\usepackage[thicklines]{cancel}
\usepackage{color}
\usepackage{CJK}
\usepackage{mathrsfs}
\usepackage{epstopdf}

\renewcommand{\theequation}{\arabic{section}.\arabic{equation}}
\renewcommand{\theequation}{\arabic{section}.\arabic{equation}}

\begin{document}
\def\a{\alpha}
\def\b{\beta}
\def\d{\delta}
\def\e{\epsilon}
\def\g{\gamma}
\def\h{\mathfrak{h}}
\def\k{\kappa}
\def\l{\lambda}
\def\o{\omega}
\def\p{\wp}
\def\r{\rho}
\def\t{\tau}
\def\s{\sigma}
\def\z{\zeta}
\def\x{\xi}
\def\V={{{\bf\rm{V}}}}
 \def\A{{\cal{A}}}
 \def\B{{\cal{B}}}
 \def\C{{\cal{C}}}
 \def\D{{\cal{D}}}
\def\K{{\cal{K}}}
\def\O{\Omega}
\def\R{\bar{R}}
\def\T{{\cal{T}}}
\def\L{\Lambda}
\def\f{E_{\tau,\eta}(sl_2)}
\def\E{E_{\tau,\eta}(sl_n)}
\def\Zb{\mathbb{Z}}
\def\Cb{\mathbb{C}}

\def\R{\overline{R}}

\def\beq{\begin{equation}}
\def\eeq{\end{equation}}
\def\bea{\begin{eqnarray}}
\def\eea{\end{eqnarray}}
\def\ba{\begin{array}}
\def\ea{\end{array}}
\def\no{\nonumber}
\def\le{\langle}
\def\re{\rangle}
\def\lt{\left}
\def\rt{\right}

\baselineskip=20pt

\title{Off-diagonal approach to the exact solution of quantum integrable systems\thanks{Project supported by the National Key R$\&$D Program of China (Grant No.2021YFA1402104), the National Natural Science Foundation of China (Grant Nos. 12247103, 12305005, 12074410, 11934015 and 11975183), Major Basic Research Program of Natural Science of Shaanxi Province (Grant Nos. 2021JCW-19 and 2017ZDJC-32), Strategic Priority Research Program of the Chinese Academy of Sciences (Grant No. XDB33000000), Young Talent Fund of Xi'an Association for Science and Technology (Grant No. 959202313086), and Shaanxi Fundamental Science Research Project for Mathematics and Physics (Grant No. 22JSZ005).}}

\author{Yi Qiao $^{1}$, \ Junpeng Cao $^{2,3,4,5}$\thanks{Corresponding author. E-mail:junpengcao@iphy.ac.cn}, \ Wen-Li Yang $^{1,5,6}$\thanks{Corresponding author. E-mail:wlyang@nwu.edu.cn}, \, \ \\ Kangjie Shi$^{1}$ \ and \ Yupeng Wang$^{2,5}$}

\date{}
\maketitle

\vspace{-0.5cm}
\begin{center}
$^{1}${Institute of Modern Physics, Northwest University, Xian 710127, China}\\  
$^{2}${Beijing National Laboratory for Condensed Matter Physics, Institute of Physics, Chinese Academy of Sciences, Beijing 100190, China}\\ 
$^{3}${School of Physical Sciences, University of Chinese Academy of Sciences, Beijing 100049, China}\\
$^{4}${Songshan Lake Materials Laboratory, Dongguan, Guangdong 523808, China}\\   
$^{5}${Peng Huanwu Center for Fundamental Theory, Xian 710127, China}\\   
$^{6}${Shaanxi Key Laboratory for Theoretical Physics Frontiers, Xian 710127, China}
\end{center}
\vspace{0.5cm}

\begin{abstract}
We investigate the $t-W$ scheme for the anti-ferromagnetic XXX spin chain under both periodic and open boundary conditions.
We propose a new parametrization of the eigenvalues of transfer matrix.
Based on it, we obtain the exact solution of the system.
By analyzing the distribution of zero roots at the ground state,
we obtain the explicit expressions of the eigenfunctions of the transfer matrix and the associated $\mathbb{W}$ operator (see (\ref{t-W-relation-op}) and (\ref{W-open})) in the thermodynamic limit.
We find that the ratio of the quantum determinant with the eigenvalue of $\mathbb{W}$ operator for the ground state exhibits exponential decay behavior. Thus this fact ensures that the so-called inversion relation (the $t-W$ relation without the $W$-term) can be used to study the ground state properties of quantum integrable systems with/without $U(1)$-symmetry in the thermodynamic limit.
\end{abstract}

\textbf{Keywords:} Quantum spin chain; Bethe ansatz; Yang-Baxter equation

\textbf{PACS:} 75.10.Pq, 02.30.Ik, 03.65.Vf 

\section{Introduction}
\setcounter{equation}{0}

Quantum integrable systems, defined by the Yang-Baxter equation \cite{yang67, bax82} or the Lax representation \cite{lax68}, provide crucial insights in quantum field theory, condensed matter physics and statistical physics. They serve as reliable benchmarks for studying many-body effects, settling debates on fundamental concepts, and exhibiting phenomena such as thermodynamic phase transitions\cite{ons44} and the generation of fractional charges\cite{fad81}. These models also have many applications in the different fields, such as cold atoms \cite{lieb63,lieb67,lieb68,gau67,guan13} and AdS/CFT correspondence \cite{bei12,chen04}.

The eigenvalue problem of quantum integrable systems with $U(1)$ symmetry has been tackled by using the methods including coordinate Bethe Ansatz \cite{bet71}, $T-Q$ relation \cite{bax71-1,bax71-2} and algebraic Bethe Ansatz \cite{tak79,skl80,fad80,skl82,tak85}. It should be emphasized that there exist the integrable models which do not possess the $U(1)$ symmetry.
Due to the $U(1)$ symmetry broken, it is very hard to construct the suitable reference state. Thus the exact solution of this kind of integrable
systems is a challenging issue. Several techniques have been developed to address this problem, including gauge transformation \cite{cao03}, fusion-based $T-Q$ relation \cite{yun95,nep021}, $q$-Onsager algebra \cite{bas1,bas2}, separation of variables \cite{nie2-2,nic13-1}, modified algebraic Bethe ansatz \cite{bel13-1,bel13-2} and off-diagonal Bethe ansatz\cite{cysw,book}.
The eigenvalues of the transfer matrix of the quantum integrable systems without $U(1)$ symmetry are characterized by the inhomogeneous $T-Q$ relations.
However, the associated Bethe ansatz equations (BAEs) are inhomogeneous and the corresponding distributions of Bethe roots are very complicated. Consequently,
the thermodynamic Bethe ansatz \cite{yang69,gau71,tak71,tak72,tak74} does not work and it is very hard to calculate the exact physical properties in the thermodynamic limit, such as the ground state,
elementary excitations and thermodynamic quantities including specific heat and magnetic susceptibility at the finite temperature.

Recently, a novel Bethe ansatz known as the $t-W$ scheme has been proposed \cite{qiao20}, which effectively tackles the challenges posed by the inhomogeneous $T-Q$ relations.
Taking the $XXZ$ spin chain with open boundary condition as an example, we show the power of this approach with the help of inhomogeneous parameters \cite{qiao21}.
The main advantage of this method is that the related BAEs are homogeneous and we can take the thermodynamic limit.
We calculated the exact physical quantities such as ground state energy, elementary excitations and surface energy.
We also extended the method to the twisted boundary situation \cite{xio21}.
Subsequently, we generalized the $t-W$ method to the finite temperature. The thermodynamic quantities including
the free energy of the XXX spin chain with periodic boundary condition are computed \cite{lu21}. Later, we have applied
this method to the supersymmetric $t-J$ model \cite{yi21}, Hubbard model \cite{yi22} and integrable $J_1-J_2$ model with competition interactions \cite{wang22,wang23}, among other notable achievements \cite{sun22,li22}.

The $t-W$ relation can be used to determine the energy spectrum directly.
Typically, the $W$ operator can be neglected in the thermodynamic limit, resulting in the $t-W$ relation becoming equivalent to the inversion relation\cite{bax82,str79,sha81}, but the exact proof is absent. Now, we focus on this issue. Our first investigation is as follows. By putting the inhomogeneous parameters into the transfer matrix,
we prove that the $t-W$ relations are closed at the inhomogeneous points, where the coefficients of $W$ terms are zero.
The inhomogeneous parameters are utilized as the auxiliary functions to determine the distribution of zero roots
in the thermodynamic limit \cite{qiao21}.
In this paper, we analytically obtain the eigenfunction of $W$ operator in the thermodynamic limit at the ground state, and
demonstrate that the leading term in the $t-W$ relation is quantum determinant instead of the $W$ term.
We take the isotropic spin-${\frac 12}$ chain as an example. The model Hamiltonian reads
\bea
H^p=\sum_{n=1}^{N}
  (\sigma_n^x\sigma_{n+1}^x+\sigma_n^y\sigma_{n+1}^y
  +\sigma_n^z\sigma_{n+1}^z),\label{PBC}
\eea
for the periodic boundary condition (PBC)
and
\bea
H^o=\sum_{n=1}^{N-1}
  (\sigma_n^x\sigma_{n+1}^x+\sigma_n^y\sigma_{n+1}^y
  +\sigma_n^z\sigma_{n+1}^z)+\frac{\eta}{p} \sigma_1^z + \frac{\eta}{q} (\sigma^z_N + \xi \sigma^x_N),\label{OBC}
\eea
for the open boundary condition (OBC), where $\sigma_n^{\alpha}$ is the Pauli
matrix along the $\alpha$-direction at $n$th site, $\alpha=(x,\,y,\,z)$,
$\sigma_{N+1}^{\alpha}=\sigma_1^{\alpha}$, $p,\, q,\, \xi$ are the boundary parameters associated with the boundary fields and $\eta=i$.

The paper is organized as follows. In section 2, we introduce the XXX spin chain with periodic boundary condition.
We show the integrability, $t-W$ solutions and the eigenfunction of transfer matrix and $W$ operator at the ground state in the thermodynamic limit.
In section 3, we generalized these results to the open boundary case. Section 4 includes the summary and further discussions.
Appendix A gives the detailed derivation of the $t-W$ relation and appendix B shows the hermitian property of transfer matrix with
PBC.

\section{Closed chain}
\setcounter{equation}{0}
\subsection{Integrability}

Throughout this paper, ${V}$ denotes a two-dimensional linear space and $\{|m\rangle, m=0,1\}$ are the orthogonal bases of it.
We shall adopt the standard notations. For any matrix $A\in {\rm End}({ V})$, $A_j$ is an
embedding operator in the tensor space ${ V}\otimes {V}\otimes\cdots$, which acts as $A$ on the $j$-th space and as identity on the other factor spaces.
For the matrix $B\in {\rm End}({ V}\otimes { V})$, $B_{i, j}$ is an embedding operator of $B$ in the tensor space, which acts as identity
on the factor spaces except for the $i$-th and $j$-th ones.

Let us introduce the $R$-matrix $R_{0,j}(u)\in {\rm End}({ V}_0\otimes { V}_j)$
\begin{equation}
  R_{0,j}(u) = \left( \begin{array}{llll}
    u + \eta & 0 & 0 & 0\\
    0 & u & \eta & 0\\
    0 & \eta & u & 0\\
    0 & 0 & 0 & u + \eta
  \end{array} \right), \label{R-matrix}
\end{equation}
where $u$ is the spectral parameter and $\eta$ is the crossing parameter.
The $R$-matrix \eqref{R-matrix} has the following properties
\bea
&&\hspace{-0.5cm}\mbox{ Initial
condition}:\,R_{0,j}(0)=  \eta P_{0,j},\no \\
&&\hspace{-0.5cm}\mbox{ Unitary
relation}:\,R_{0,j}(u)R_{j,0}(-u)= \phi(u)\times {\rm id},\no \\
&&\hspace{-0.5cm}\mbox{ Crossing
relation}:\,R_{0,j}(u)=-\sigma_0^y R_{0,j}^{t_0}(-u-\eta)\sigma_0^y,\no \\
&&\hspace{-0.5cm}\mbox{ PT-symmetry}:\,R_{0,j}(u)=R_{j,0}(u)=R_{0,j}^{t_0\,t_j}(u),\no \\
&&\hspace{-0.5cm}\mbox{ $Z_2$-symmetry}:\,\sigma_0^\alpha \sigma_j^\alpha R_{0,j}(u)= R_{0,j}(u) \sigma_0^\alpha \sigma_j^\alpha, \mbox{ for } \alpha=x,y,z,\no \\
&&\hspace{-0.5cm}\mbox{ Fusion condition}:\,R_{0,j}(\pm\eta)=\eta (\pm 1+P_{0,j})=\pm2\eta P_{0,j}^{(\pm)},\label{XXX-R-property}
\eea
where $\phi(u)=\eta^2-u^2$, $t_0$ (or $t_j$) denotes the
transposition in the space ${V}_0$ (or ${V}_j$),
$P_{0,j}$ is the permutation operator possessing the property
\begin{eqnarray}
  R_{j,k}(u)=P_{0,j}R_{0,k}(u)P_{0,j},
\end{eqnarray}
$P^{(-)}_{0,j}$ is the one-dimensional antisymmetric project operator defined in the one-dimensional subspace spanned by
$\frac{1}{\sqrt{2}}(|12\rangle_{0,j}-|21\rangle_{0,j})$,
$P^{(+)}_{0,j}$ is the three-dimensional symmetric projector defined in the three-dimensional subspace
spanned by the orthogonal bases $\{|11\rangle_{0,j},\,\frac{1}{\sqrt{2}}(|12\rangle_{0,j} +|21\rangle_{0,j}),\,|22\rangle_{0,j}\}$,
$P^{(-)}_{0,j}+P^{(+)}_{0,j}={\rm id}$. The $R$-matrix (\ref{R-matrix}) satisfies the Yang-Baxter equation (YBE)
\bea
&&R_{1,2}(u_1-u_2)R_{1,3}(u_1-u_3)R_{2,3}(u_2-u_3)
\no \\ && \quad =R_{2,3}(u_2-u_3)R_{1,3}(u_1-u_3)R_{1,2}(u_1-u_2).\label{QYB}
\eea
Starting from the $R$-matrix, we construct the monodromy matrix as
\bea\label{monodromy-matrix}
T_0(u)=R_{0,N}(u-\theta_N) R_{0,N-1}(u-\theta_{N-1}) \cdots R_{0,1}(u-\theta_1),
\eea
where $V_0$ is the auxiliary space, $V_1\otimes V_2 \otimes \cdots \otimes V_{N}$ is the physical or quantum space,
$N$ is the number of sites and $\{\theta_j|j=1, \cdots, N\}$ are the inhomogeneous parameters.
The transfer matrix $t(u)$ of the XXX spin-1/2 closed chain is given by
 \begin{eqnarray}
 t(u)=tr_0 T_0(u),\label{trans-per}
 \end{eqnarray}
where $tr_0$ denotes trace over the auxiliary space $V_0$.
The transfer matrix $t(u)$ is the generating function of conserved quantities in the system, and the
Hamiltonian \eqref{PBC} is generated by the $t(u)$ as
\begin{equation}\label{Ham-def-p}
  H^p = 2 \eta \left. \frac{\partial \ln t (u)}{\partial u} \right|_{u = 0,
  \{\theta_j = 0\}} - N.
\end{equation}
The transfer matrices with different spectral parameters commute mutually, i.e., $[t (u), t (v)] = 0$, which ensures the integrability of the model \eqref{PBC}.

\subsection{$t-W$ scheme}

By using the fusion technique \cite{kul81,kir86}, we consider the product of transfer matrices $t(u)$ and $t(u-\eta)$
\bea
t(u)t(u-\eta)&=&tr_{1,2}\lt\{T_2(u)\,T_1(u-\eta)\rt\}\no\\[4pt]
&=&tr_{1,2}\lt\{T_2(u)\,T_1(u-\eta)(P^{(-)}_{1,2}+P^{(+)}_{1,2})\rt\}\no\\[4pt]
&=&tr_{1,2}\lt\{P^{(-)}_{1,2}T_2(u)\,T_1(u-\eta)P^{(-)}_{1,2}\rt\}+tr_{1,2}\lt\{P^{(+)}_{1,2}T_2(u)\,T_1(u-\eta)P^{(+)}_{1,2}\rt\} \no \\
&=&a(u)\,d(u-\eta)\times {\rm id}+d(u)\mathbb{W}(u),\label{t-W-relation-op}
\eea
where the functions $a(u)$ and $d(u)$ are given by
\begin{eqnarray}
a(u)=\prod_{j=1}^N(u-\theta_j+\eta),\quad d(u)=a(u-\eta),\label{a-d-functions}
\end{eqnarray}
$a(u)d(u-\eta)$ is the quantum determinant and $\mathbb{W}(u)$ is a new operator.
The detailed proof is given in Appendix A.
At the points of $\{u=\theta_j\}$, the operator relation \eqref{t-W-relation-op} can be simplified as
\begin{eqnarray}
&&t(\theta_j)t(\theta_j-\eta) =a(\theta_j)\,
d(\theta_j-\eta),\quad j=1,\cdots,N, \label{Eigen-idenstity}
\end{eqnarray}
due to the fact $d(\theta_j)=0$.

From the definition, we know that both
$t(u)$ and $\mathbb{W}(u)$ are the operator-valued
polynomial of $u$ with degree $N$. Moreover, the matrices $t(u)$ and $\mathbb{W}(u)$ commutate with each other, namely,
\bea
[t(u),\,t(v)]=[\mathbb{W}(u),\,\mathbb{W}(v)]=[t(u),\,\mathbb{W}(v)]=0.\label{Communtivity}
\eea
Thus they have common eigenstates.
Acting the operator identities \eqref{t-W-relation-op} and \eqref{Eigen-idenstity}
on the common eigenstate $|\Psi\rangle$, we obtain the $t-W$ relation
\bea
&&\Lambda(u)\,\Lambda(u-\eta)=a(u)\,d(u-\eta)+d(u) W(u),\label{t-W-relation-Eigen}\\
&&\Lambda(\theta_j)\Lambda(\theta_j-\eta) =a(\theta_j)\,
d(\theta_j-\eta),\quad j=1,\cdots, N, \label{Eigqwen-identity}
\eea
where $\Lambda(u)$ and $W(u)$ are the eigenvalues of the transfer matrix $t(u)$ and $\mathbb{W}(u)$ operator, respectively
\bea
t(u)\,|\Psi\rangle=\Lambda(u)\,|\Psi\rangle,\quad \mathbb{W}(u)\,|\Psi\rangle=W(u)\,|\Psi\rangle.\no
\eea
From the construction of transfer matrix (\ref{trans-per}), we conclude that
\bea
&&\mbox{$\Lambda(u)$, as a function of $u$, is a  polynomial of degree $N$}, \label{Elliptic-Poly-1}\\
&&\mbox{$W(u)$, as a function of $u$, is a  polynomial of degree $N$}.\label{Elliptic-Poly-2}
\eea
Meanwhile, $\Lambda(u)$ and $W(u)$ satisfy the asymptotic behaviors
\bea
\lim_{u\rightarrow \infty} \Lambda(u)=2u^N+\cdots,\quad \lim_{u\rightarrow \infty} {W}(u)=3u^N+\cdots. \label{Spin-op}
\eea

\subsection{Exact solution}
\label{XXZclosedchain}

In order to obtain the exact solution of the spin-$\frac{1}{2}$ XXX closed chain described by the Hamiltonian (\ref{PBC}),
let us take the homogeneous limit, i.e., $\{\theta_j= 0\}$.
Usually, the eigenvalues $\Lambda(u)$ and $W(u)$ are expressed by the $T-Q$ relations with the help of Bethe roots. Here, we quantify
the $\Lambda(u)$ and $W(u)$ by their zero roots as
\bea
\Lambda(u)= 2\prod_{j=1}^{N} (u-z_j+\frac{\eta}{2}), \quad
W(u)=3 \prod_{j=1}^N (u-w_j),\label{Expansion-4}
\eea
where $\{z_j|j=1,\cdots,N\}$ and $\{w_j|j=1,\cdots,N\}$ are the zero roots of $\Lambda(u)$ and $W(u)$, respectively.

Substituting Eq.\eqref{Expansion-4} with $\{u=z_j-\frac{\eta}{2}\}$ and $\{u=w_j\}$ into
into the $t-W$ relation (\ref{t-W-relation-Eigen}) with $\{\theta_j=0\}$, we obtain that the zero roots $\{z_j\}$ and
$\{w_j\}$ should satisfy the BAEs
\bea
&&(z_j+\frac{\eta}{2})^N\,(z_j-\frac{3}{2}\eta)^N =-(z_j-\frac{\eta}{2})^N\,W(z_j-\frac{\eta}{2}),\quad j=1,\cdots,N,\label{BAE-1}\\[4pt]
&&\Lambda(w_j)\,\Lambda(w_j-\eta)=(w_j+\eta)^N\,(w_j-\eta)^N, \quad j=1,\cdots,N.\label{BAE-2}
\eea
Further analysis gives that the $2N$ roots satisfy the constrains
\bea
\{z^*_j\}=\{z_j\},\quad \{w^*_j\}=\{w_j\}.\label{Zero-Constrain}
\eea
The detailed proof can be found in Appendix B.

The eigenvalues of the Hamiltonian (\ref{PBC}) can be expressed in terms of the zero roots $\{z_j\}$ as
\bea
E^p=-2\eta\times\sum_{j=1}^{N}\frac{1}{z_j-\frac{\eta}{2}} -N.\label{Energy}
\eea
For the finite system size $N$, one can solve the BAEs \eqref{BAE-1} and \eqref{BAE-2} numerically. Substituting
the values of roots into Eq.\eqref{Energy}, one obtains the energy of the system.
The most interesting thing is the thermodynamic limit where $N$ tends to infinity, which will be addressed in the next subsection.

\subsection{Ground state eigenfunctions in the thermodynamic limit}
\begin{figure}[htbp]
\begin{center}
\includegraphics[scale=0.8]{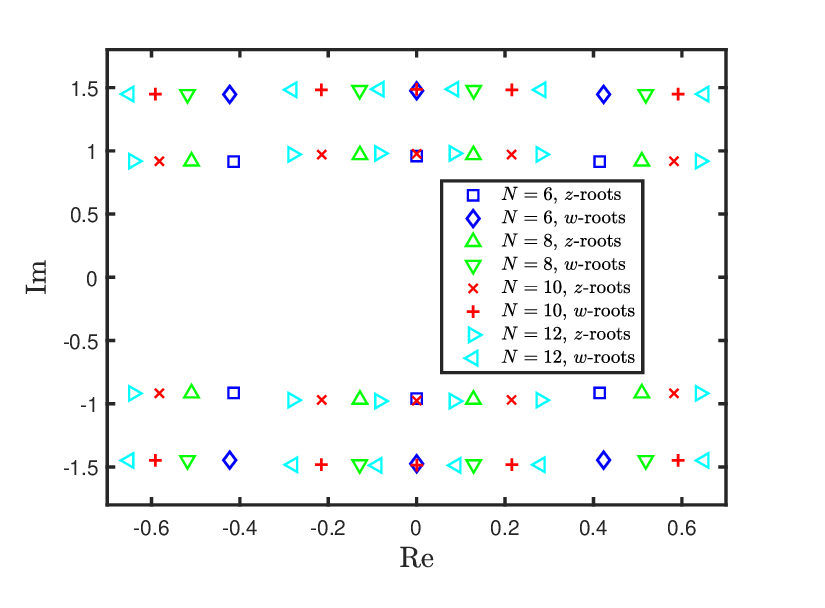}\\[4pt]
\parbox[c]{15.0cm}{\footnotesize{\bf Fig.~1.} Patterns of zero roots at the ground state with $N=6,8,10,12$. The data are obtained by using the exact numerical diagonalization with the inhomogeneous parameters $\{{\theta}_j=0\}$.}
\end{center}
\end{figure}
We first consider the distribution of zero roots. According to the exact numerical calculation with finite $N$ and
the analytic analysis with infinite $N$, we find that
all the zero roots form the conjugate pairs around the line $\pm\frac{l\eta}{2}$ with a positive integer $l \geq 2$, which is called as the $l$-strings.
At the ground state, both $z$-roots and $w$-roots form the 2-strings. For even $N$, the patterns of roots with $N=6,8,10,12$ are shown in figure 1. We can find that all the $z$-roots form conjugate pairs as $\{u_j^{(2)}\pm \eta|j=1,\cdots,N/2\}$ with real $u_j^{(2)}$ and all the $w$-roots form conjugate pairs as $\{\bar{u}_j^{(2)}\pm \frac{3\eta}2 |j=1,\cdots,N/2\}$ with real $\bar{u}_j^{(2)}$.
Substituting the $2$-strings into Eq.\eqref{Expansion-4}, we obtain the ground state eigenvalues $\Lambda_g(u)$ and $W_g(u)$ as
\bea
&&\Lambda_g(u)= 2
\prod_{j=1}^{N/2}(u-u_j^{(2)}-\frac{\eta}{2})(u-u_j^{(2)}+\frac{3\eta}{2}), \label{t-gs-p}\\[4pt]
&&W_g(u)=3
\prod_{j=1}^{N/2} (u-\bar{u}_j^{(2)}-\frac32 \eta)(u-\bar{u}_j^{(2)}+\frac32 \eta).\label{w-gs-p}
\eea
Now we analyze the leading terms of the eigenvalues.
For this purpose, we define
\bea
&&\Lambda_g(u)=e^{N[ \lambda_g^{(0)}(u)+\frac1N \lambda_g^{(1)}(u)+O(\frac1{N^2}) ]}, \label{t-gs-p1}\\[4pt]
&&W_g(u)=e^{N[ \omega_g^{(0)}(u)+\frac1N \omega_g^{(1)}(u)+O(\frac1{N^2}) ]},\label{w-gs-p1}
\eea
where $\lambda_g^{(0)}(u)$ and $\omega_g^{(0)}(u)$ are the highest order terms of $u$,
and $\lambda_g^{(1)}(u)$ and $\omega_g^{(1)}(u)$ are the second higher order terms.
In the thermodynamic limit $N\to\infty$, the leading terms of $\Lambda_g(u)$ can be obtained by
taking the derivatives of logarithm of Eqs.\eqref{t-gs-p} and \eqref{t-gs-p1}
\bea
\frac{\partial}{\partial u}\lambda_g^{(\beta)}(u)=\int_{-\infty}^{\infty} (\frac{1}{u-\lambda-\frac\eta2}+\frac{1}{u-\lambda+\frac{3\eta}2}) \rho^{(\beta)}_g(\lambda)\mathrm{d}\lambda, \quad \beta=0,1, \label{lamg-o}
\eea
where $\rho^{(\beta)}_g(\lambda)$ is the density of $z$-roots at the ground state. The leading terms of $W_g(u)$ are determined by
taking the derivatives of logarithm of Eqs.\eqref{w-gs-p} and \eqref{w-gs-p1} as
\bea
\frac{\partial}{\partial u}w_g^{(\beta)}(u)=\int_{-\infty}^{\infty} (\frac{1}{u-\lambda-\frac{3\eta}2}+\frac{1}{u-\lambda+\frac{3\eta}2}) \rho^{(\beta)}_w(\lambda)\mathrm{d}\lambda \quad \beta=0,1, \label{laxmg-o}
\eea
where $\rho^{(\beta)}_w(\lambda)$ is the density of $w$-roots.

The role of inset inhomogeneous parameters is to help us to determine the density of $z$-roots. Take the difference
of Eq.\eqref{Eigqwen-identity} at two nearest inhomogeneous points. In the thermodynamic limit,
we set that the density of inhomogeneous parameters as the $\delta$-function. Then we have
\bea
\frac{\partial}{\partial u}\ln[\Lambda_g(u)\Lambda_g(u-\eta)]=\frac{\partial}{\partial u} \ln[(u+\eta)^N(u-\eta)^N]+O(\frac1N),\label{lamlam1}
\eea
where $O(\frac1N)$ is the order-dependent correction.
Substituting \eqref{t-gs-p1} into \eqref{lamlam1}, we have
\bea\label{lamlam2}
\frac{\partial}{\partial u}\big[\lambda_g^{(0)}(u)+\lambda_g^{(0)} (u-\eta)\big]&=&\frac{1}{u+\eta}+\frac{1}{u-\eta},\qquad \lambda_g^{(0)} (0)=0, \no \\
\frac{\partial}{\partial u}\big[\lambda_g^{(1)}(u)+\lambda_g^{(1)} (u-\eta)\big]&=&0, \qquad \lambda_g^{(1)} (0)=0.
\eea
Substituting Eq.\eqref{lamg-o} with $\beta=0$ into \eqref{lamlam2}, we have
\bea
\int_{-\infty}^{\infty} \big(\frac{1}{u-\lambda-\frac\eta2}+\frac{1}{u-\lambda+\frac{\eta}2} +\frac{1}{u-\lambda-\frac{3\eta}2}+\frac{1}{u-\lambda+\frac{3\eta}2}\big) \rho^{(0)}_g(\lambda)\mathrm{d}\lambda=\frac{1}{u+\eta}+\frac{1}{u-\eta}.\label{oop}
\eea
Solving Eq.\eqref{oop} by the Fourier transformation
\bea
\tilde{f}(\omega)=\frac1{2\pi}\int^{\infty}_{-\infty}f(t)e^{-i\omega t}\mathrm{d}t, \quad
f(t)=\int^{\infty}_{-\infty}\tilde{f}(\omega)e^{i\omega t}\mathrm{d}w,
\eea
we obtain the solution of densities of $z$-roots as
\bea
\rho^{(0)}_g(\lambda)=\frac1{2\cosh(\pi\lambda)}, \qquad \rho^{(1)}_g(\lambda)=0.\label{lam0g-o}
\eea
By incorporating the 2-string structure into the energy expression \eqref{Energy} and integrating it with the densities \eqref{lam0g-o} as the weighting factor, we can directly determine the ground state energy in the thermodynamic limit
\bea
E_g&=&-2N i \int_{-\infty}^{\infty}\bigg(\frac1{\lambda+\frac i2}+\frac1{\lambda-\frac {3i}2} \bigg) \big(\rho^{(0)}_w(\lambda)+ \rho^{(1)}_w(\lambda)\big) \, \mathrm{d} \lambda- N \no \\ [4pt]
&=&(1-4\ln2)N,
\eea
which is consistent with the result obtained through algebraic Bethe ansatz\cite{book}.
Substituting Eq.\eqref{lam0g-o} into \eqref{lamg-o} and taking the integral, we obtain
\bea\label{lam0-p}
\lambda_g^{(0)}(u)=\ln\frac{2\Gamma(1+\frac{iu}2) \Gamma(\frac32-\frac{iu}2)}{\Gamma(\frac12+\frac{iu}2) \Gamma(1-\frac{iu}2)}, \quad \lambda_g^{(1)}(u)=0.\label{la1m0g-o}
\eea
In the derivation, we have used the formula
\bea
\int_{0}^{\infty}\frac{e^{-t}-e^{-ut}}{1-e^{-t}}\mathrm{d}t-\gamma =\frac{\mathrm{d}}{\mathrm{d}u}\ln\Gamma(u),
\eea
where $\gamma$ is the Euler-Mascheroni constant. According to the formula
\bea
\ln\Gamma(x+1)=\frac12\ln(2\pi x)+x\ln(x)-x+\ln(1+\frac1{12 x}+\frac1{288 x^2}+\cdots),
\eea
the $\lambda_g^{(0)}(u)$ can be expanded with respect to $u$ as
\bea
\lambda_g^{(0)}(u)=\ln u+\ln\frac{ \cosh{\frac{\pi u}2} } { \sinh{\frac{\pi u}2} }+\frac{\eta}{2 u}+\frac{1}{2u^2}+O(\frac1{u^2}), \quad \mathrm{Im}(u)<1. \label{oopp}
\eea
From Eqs.\eqref{t-gs-p1}, \eqref{la1m0g-o} and \eqref{oopp}, we obtain that the eigenvalue of
transfer matrix at the ground state with the thermodynamic limit is
\bea
\Lambda_g(u)=\bigg(\frac{2\Gamma(1+\frac{iu}2) \Gamma(\frac32-\frac{iu}2)}{\Gamma(\frac12+\frac{iu}2) \Gamma(1-\frac{iu}2)}\bigg)^N e^{O(\frac1N)}. \label{2}
\eea

Now, we calculate the leading terms of $W(u)$. Substituting $z_j=u_j^{(2)}+\frac\eta2$ into the BAEs (\ref{BAE-1}), we obtain
\bea
(u_j^{(2)}+\frac{3\eta}{2})^N\, (u_j^{(2)}-\frac{\eta}{2})^N = -(u_j^{(2)}+\frac{\eta}{2})^N\,W_g(u_j^{(2)}+\frac{\eta}{2}),\quad j=1,\cdots, \frac{N}2,\label{BAE-1-g-q}
\eea
where $\{u_j^{(2)}\}$ are real.
Taking the complex conjugation of above relation, we have
\bea
(u_j^{(2)}+\frac\eta2)^N\,(u_j^{(2)}-\frac{3\eta}2)^N =-(u_j^{(2)}-\frac{\eta}2)^N W_g(u_j^{(2)}-\frac\eta2),\quad j=1,\cdots, \frac{N}2.\label{BAE-2-g-q}
\eea
Multiplying Eq.\eqref{BAE-1-g-q} with \eqref{BAE-2-g-q}, we obtain
\bea
(u_j^{(2)}+\frac{3\eta}{2})^N(u_j^{(2)}-\frac{3\eta}{2})^N = W_g(u_j^{(2)}+\frac\eta2)W_g(u_j^{(2)}-\frac\eta2),\quad j=1,\cdots, \frac{N}2.
\eea
Taking the derivative of logarithm of above equation, we arrive at
\bea\label{ww1}
\frac1N \frac{\partial}{\partial u} \ln [W_g(u+\frac\eta2) W_g(u-\frac\eta2)]=\frac1{u+\frac{3\eta}2} +\frac1{u-\frac{3\eta}2}+O(\frac1{N^2}).
\eea
Substituting Eq.\eqref{w-gs-p1} into \eqref{ww1}, we have
\bea\label{ww2}
\frac{\partial}{\partial u}[w_g^{(0)}(u+\frac\eta2)+w_g^{(0)} (u-\frac\eta2)]&=&\frac{1}{u+\frac{3\eta}{2}}+\frac{1}{u-\frac{3\eta}{2}}, \no \\
\frac{\partial}{\partial u}[w_g^{(1)}(u+\frac\eta2)+w_g^{(1)} (u-\frac\eta2)]&=&0.
\eea
Substituting Eq.\eqref{laxmg-o} into above equations and solving it by the Fourier transformation, we obtain the density of $w$-roots as
\bea
\rho^{(0)}_w(\lambda)=\frac1{2\cosh(\pi\lambda)}, \quad \rho^{(1)}_w(\lambda)=0. \label{wBAE-2-g-q}
\eea
Substituting Eq.\eqref{wBAE-2-g-q} into \eqref{laxmg-o}, we have
\bea\label{w0-p}
\frac{\partial}{\partial u} w_g^{(0)}(u) =\frac{\partial}{\partial u} \ln\frac{\Gamma(\frac32+\frac{iu}2) \Gamma(\frac32-\frac{iu}2)}{\Gamma(1+\frac{iu}2) \Gamma(1-\frac{iu}2)}
=\frac{\partial}{\partial u} \ln \bigg(\frac{(u+\eta)(u-\eta)}{2u} \tanh\frac{\pi u}2\bigg).
\eea
Taking the integration, we obtain
\bea\label{w01-p}
w_g^{(0)}(u) = \ln\bigg( \frac{C_w^{(0)}}{2}\frac{(u+\eta)(u-\eta)}{u} \tanh\frac{\pi u}2\bigg),
\eea
where $C_w^{(0)}$ is the integration constant. Thus the ground state eigenvalue $W_g(u)$ can be expressed as
\bea\label{wg-p}
W_g(u) = \bigg(\frac{(u+\eta)(u-\eta)}{u}\bigg)^N \bigg( \frac{C_w^{(0)}}{2}\tanh\frac{\pi u}2\bigg)^N e^{\omega_g^{(1)}(0)}e^{O(\frac1N)}.
\eea

Substituting Eq.\eqref{wg-p} into the $t-W$ relation \eqref{t-W-relation-Eigen} with $\{\theta_j=0\}$, we obtain
\bea\label{tw1}
\Lambda_g(u)\Lambda_g(u-\eta)=(u+\eta)^N(u-\eta)^Ne^{O(\frac1N)}
=(u+\eta)^N(u-\eta)^N\bigg[ 1 + \bigg(\frac{C_w^{(0)}}{2}\tanh\frac{\pi u}2\bigg)^N C_w^{(1)}e^{O(\frac1N)}\bigg].\label{lamglamg-p}
\eea
When $u\to \infty$, from the asymptotic behavior of $\Lambda_g(u)$ or Eq.\eqref{2}, we know that the coefficient of highest order term of left hand side of Eq.\eqref{lamglamg-p} is 4.
In order to meets this constraint, the left hand side of Eq.\eqref{lamglamg-p} should also be 4, which gives that $C_w^{(0)}=2$ and $C_w^{(1)}=3$.
Due to the fact that $\tanh\frac{\pi u}2<1$, the second term of the right hand side of \eqref{tw1} turns to zero when $N\to\infty$,
which gives that the $W$ function can be neglected in the thermodynamic limit.
Then we conclude that the $t-W$ relation \eqref{t-W-relation-Eigen} can be used to study the
ground state physical properties.

\section{Open chain}
\setcounter{equation}{0}
\subsection{Integrability}

Next, we consider the open boundary condition. The boundary reflections are characterized by the reflection matrices
\begin{eqnarray}
  K^- (u) = \left( \begin{array}{ll}
    p + u & 0\\
    0 & p - u
  \end{array} \right), \quad
  K^+ (u)  = \left( \begin{array}{ll}
    q + u + \eta & \xi (u + \eta)\\
    \xi (u + \eta) & q - u - \eta
  \end{array} \right),
\end{eqnarray}
which satisfy the reflection equation (RE)
\begin{eqnarray}\label{REle}
R_{1,2}(\lambda-u)K_{1}^{-}(\lambda)R_{2,1}(\lambda+u) K_{2}^{-}(u)=K_{2}^{-}(u)R_{1,2}(\lambda+u)K_{1}^{-}(\lambda) R_{2,1}(\lambda-u),
\end{eqnarray}
and the dual reflection equation
\begin{eqnarray}\label{REre}
&&R_{1,2}(-\lambda+u)K_{1}^{+}(\lambda)R_{2,1}(-\lambda-u-2\eta) K_{2}^{+}(u) \nonumber\\
&&=K_{2}^{+}(u)R_{1,2}(-\lambda-u-2\eta)K_{1}^{+}(\lambda) R_{2,1}(-\lambda+u).
\end{eqnarray}
Due to the boundary reflection, we should introduce the reflecting monodromy matrix
\bea\label{monodromy-matrix-hat}
\hat{T}_0(u)=R_{0,1}(u+\theta_1) R_{0,2}(u+\theta_2) \cdots R_{0,N}(u+\theta_N).
\eea
The double-row monodromy matrix $\mathscr{U}_0(u)$ is
\bea\label{double-row-transfer-XXX-open}
\mathscr{U}_0(u)=T_0(u)K^-_0(u)\hat{T}_0(u),
\eea
which satisfies the RE
\bea
R_{1,2}(\lambda-u)\mathscr{U}_1(\lambda)R_{2,1}(\lambda+u)\mathscr{U}_2(u) =\mathscr{U}_2(u)R_{1,2}(\lambda+u)\mathscr{U}_1(\lambda)R_{2,1}(\lambda-u).
\eea
The transfer matrix for the open boundary case is constructed as
\bea\label{transfer-XXX-open}
t^o(u) = tr_0\{K^+_0(u) \mathscr{U}_0(u)\}.
\eea
Using the crossing symmetry of the R-matrix (\ref{XXX-R-property}), we can demonstrate
\bea\label{tt-r}
t^o(u)=t^o(-u-\eta).
\eea
The YBE and RE lead to that the transfer matrices with different spectral parameters commutate mutually, i.e., $[t^o (u), t^o (v)] = 0$.
Thus $t^o(u)$ is the generating function of conserved quantities and the system is integrable.
The Hamiltonian \eqref{Ham-def-o} is generated by the transfer matrix $t(u)$ as
\begin{equation}\label{Ham-def-o}
  H^o = \eta \left. \frac{\partial \ln t^o (u)}{\partial u} \right|_{u = 0,  \{\theta_j = 0\}} - N.
\end{equation}

\subsection{$t-W$ scheme}

Following the idea of fusion, we still consider the product of two transfer matrices with certain shift of the spectral parameter
\bea
&&t^o(u)t^o(u-\eta)=tr_{1,2}\lt\{K^+_2(u)\mathscr{U}_2(u)\,K^+_1(u-\eta)\mathscr{U}_1(u-\eta)\rt\}\no\\[4pt]
&=&tr_{1,2}\lt\{K^{+t_2}_2(u)K^+_1(u-\eta)\,\mathscr{U}^{t_2}_2(u)\mathscr{U}_1(u-\eta)\rt\} \no\\[4pt]
&\stackrel{(\ref{XXX-R-property})}{=}&\frac{1}{\rho_2(2u-\eta)}
tr_{1,2}\lt\{K^{+t_2}_2(u)K^+_1(u-\eta)R_{2,1}^{t_2}(-2u-\eta)\,R_{1,2}^{t_2}(2u-\eta)\mathscr{U}^{t_2}_2(u)\mathscr{U}_1(u-\eta)\rt\}\no\\[4pt]
&=&\frac{1}{\rho_2(2u-\eta)}
tr_{1,2}\lt\{\lt(K^+_1(u-\eta)R_{2,1}(-2u-\eta)K^+_2(u)\rt)^{t_2}\rt.\no\\[4pt]
&&\qquad\qquad\times\lt.\lt(\mathscr{U}_2(u)R_{1,2}(2u-\eta)\mathscr{U}_1(u-\eta)\rt)^{t_2}\rt\}\no\\[4pt]
&=&\frac{1}{\rho_2(2u-\eta)}
tr_{1,2}\lt\{\lt(K^+_1(u-\eta)R_{2,1}(-2u-\eta)K^+_2(u)\rt)\rt.\no\\[4pt]
&&\qquad\qquad\times\lt.\lt(\mathscr{U}_2(u)R_{1,2}(2u-\eta)\mathscr{U}_1(u-\eta)\rt)\,\lt(P^{(-)}_{2,1}+P^{(+)}_{2,1}\rt)\rt\}\no\\
&=&
\frac{1}{\rho_2(2u-\eta)}
tr_{1,2}\lt\{\lt(K^+_1(u-\eta)R_{2,1}(-\hspace{-0.06truecm}2u-\eta)
K^+_2(u)\rt)\lt(\mathscr{U}_2(u)R_{1,2}(2u-\eta)\mathscr{U}_1(u-\eta)\rt)P^{(-)}_{2,1}\rt. \no\\[4pt]
&&+
\lt.\lt(K^+_1(u-\eta)R_{2,1}(-\hspace{-0.06truecm}2u-\eta)
K^+_2(u)\rt)\lt(\mathscr{U}_2(u)R_{1,2}(2u-\eta)\mathscr{U}_1(u-\eta)\rt)P^{(+)}_{2,1}\rt\} \no\\
&=& \Delta^o(u)/\big((u+\frac\eta2)(u-\frac\eta2)\big)+t^o_2(u), \label{22}
\eea
where $\rho_2(u)=-u(u+2\eta)$.

The first term of Eq.\eqref{22} give a number which is the quantum determinant
\bea
\Delta^o(u)=a^o(u)d^o(u-\eta)(u+\frac\eta2)(u-\frac\eta2),
\eea
where the functions $a^o(u)$ and $d^o(u)$ are
\bea
a^o(u)&=&\frac{u+\eta}{u+\frac{\eta}{2}}(u+p)(\sqrt{1+\xi^2}\,u+q) \prod_{j=1}^{N}(u-\theta_j+\eta)(u+\theta_j+\eta),\\[4pt]
d^o(u)&=&a^o(-u-\eta)=\frac{u}{u+\frac{\eta}{2}}(u-p+\eta)(\sqrt{1+\xi^2}(u+\eta)-q) \prod_{j=1}^{N}(u-\theta_j)(u+\theta_j).
\eea

The second term of Eq.\eqref{22} is a new operator which is the fused transfer matrix \cite{book, cao14, hao14} up to a constant
\bea
t^o_2(u)=\frac{1}{\rho_2(2u-\eta)}
tr_{1,2}\lt\{K^{+}_{\{1,2\}}(u)T_{\{1,2\}}(u)K^{-}_{\{1,2\}}(u)\hat{T}_{\{1,2\}}(u)\rt\}.
\eea
From the fusion of the reflection matrices and mondromy matrices
\bea
&&K^{+}_{\{1,2\}}(u)= P^{(+)}_{2,1}K^+_1(u-\eta)R_{2,1}(-2u-\eta)K^+_2(u)P^{(+)}_{1,2}= 2u\,K_{\{1,2\}}^{(1)+}(u),\label{Fus-K-1}\\[4pt]
&&K^{-}_{\{1,2\}}(u)= P^{(+)}_{1,2}K^-_2(u)R_{1,2}(2u-\eta)K^-_1(u-\eta)P^{(+)}_{2,1}= 2u\,K_{\{1,2\}}^{(1)-}(u),\label{Fus-K-2}\\[4pt]
&&T_{\{1,2\}}(u)=P^{(+)}_{1,2}\,T_2(u)\,T_1(u-\eta)\,P^{(+)}_{1,2}=\prod_{l=1}^N(u-\theta_l) T_{\{1,2\}}^{(1,\frac{1}{2})}(u),\label{Fus-T-1}\\[4pt]
&&\hat{T}_{\{1,2\}}(u)=P^{(+)}_{2,1}\,\hat{T}_2(u)\,\hat{T}_1(u-\eta)\,P^{(+)}_{2,1}=\prod_{l=1}^N(u+\theta_l) \hat{T}_{\{1,2\}}^{(1,\frac{1}{2})}(u),\label{Fus-T-2}
\eea
we obtain
\bea
t^o_2(u)=\frac{4u^2}{\rho_2(2u-\eta)} d^o(u) \mathbb{W}^o(u),
\eea
where
\bea
\mathbb{W}^o(u)=tr_{\{1,2\}}\lt\{K^{(1)+}_{\{1,2\}}(u)T_{\{1,2\}}^{(1,\frac{1}{2})}(u)K^{(1)-}_{\{1,2\}}(u)\hat{T}^{(1,\frac{1}{2})}_{\{1,2\}}(u)\rt\}.\label{W-open}
\eea
Then the operator identity \eqref{22} can be expressed as
\bea
t^o(u)t^o(u-\eta)=\Delta^o(u)\times {\rm id}/((u+\frac\eta2)(u-\frac\eta2))+\frac{4u^2}{\rho_2(2u-\eta)}
d^o(u)\mathbb{W}^o(u).\label{A.211}\eea
At the inhomogeneous points $\{u=\theta_j\}$, Eq.\eqref{A.211} reduces to
\bea\label{t-theta-oo}
(\theta_l+\frac{\eta}{2})\,(\theta_l-\frac{\eta}{2}) t^o(\theta_l)t^o(\theta_l-\eta)=\Delta^o(\theta_l), \quad j=1,\cdots, N.
\eea

The fusion does not break the integrability of the system, thus the transfer matrix and the fused transfer matrix commutate with each other. Thus they have common
eigenstates. Acting the operator relation \eqref{A.211} on a common eigenstate, we obtain the $t-W$ relation
\bea\label{tw-o}
\Delta^o(u)-(u+\frac{\eta}{2})\,(u-\frac{\eta}{2})\bar{\Lambda}(u)\bar{\Lambda}(u-\eta) =u^{2}\prod_{j=1}^{N}(u-\theta_j)(u+\theta_j)\,\bar{W}(u),\label{Ration-Main-relation-1}
\eea
where $\bar{\Lambda}(u)$ and $\bar{W}(u)$ are the eigenvalues of the transfer matrix $t^o(u)$ and the fused one $\mathbb{W}^o(u)$, respectively.
At the points of inhomogeneous point, the eigenvalue $\bar \Lambda(u)$ satisfies
\bea\label{t-theta-o}
(\theta_l+\frac{\eta}{2})\,(\theta_l-\frac{\eta}{2}) \bar{\Lambda}(\theta_l)\bar{\Lambda}(\theta_l-\eta)=\Delta^o(\theta_l), \quad j=1,\cdots, N.
\eea

\subsection{Exact solution}

The exact solution of the system does not depend on the inhomogeneous parameters. Thus we set them as zero. From the definitions, we know that the eigenvalue function $\bar{\Lambda}(u)$ is a polynomial of $u$ with degree $2N+2$ and also satisfies the crossing symmetry and asymptotic behavior
\bea
\bar{\Lambda}(-u-\eta)=\bar{\Lambda}(u),\quad  \lim_{u\rightarrow \infty}\bar{\Lambda}(u)=2u^{2N+2}+\cdots.\label{Ration-Eigen-Asy-1}
\eea
The eigenvalue function $\bar{W}(u)$ is an polynomial of $u$ with degree of $2N+4$ and has asymptotic behavior
\bea
\bar{W}(-u)=\bar{W}(u),\quad  \lim_{u\rightarrow \infty} \bar{W}(u)=(\xi^2-3)u^{2N+4}+\cdots. \label{Ration-Eigen-Asy-2}
\eea
Based on above analysis, we parameterized $\bar{\Lambda}(u)$ and $\bar{W}(u)$ as
\bea
\bar{\Lambda}(u)&=&2\prod_{j=1}^{N+1}(u-z_j+\frac{\eta}{2})(u+z_j+\frac{\eta}{2}),\label{Ration-Expandions}\\ \bar{W}(u)&=&(\xi^2-3)\prod_{k=1}^{N+2}(u-w_k)(u+w_k),\label{Ration-Expandions-1}
\eea
where $\{z_j|j=1,\cdots,N+1\}$ and $\{w_k|k=1,\cdots,N+2\}$ are the roots of corresponding polynomials, which are completely determined by the BAEs
\bea
&&\Delta(z_j-\frac{\eta}{2})=(z_j-\frac{\eta}{2})^{2N+2}\,\bar{W}(z_j-\frac{\eta}{2}),\quad j=1,\cdots,N+1,\label{Rationa-BAE-1}\\[4pt]
&&\Delta(w_k)=(w_k+\frac{\eta}{2})(w_k-\frac{\eta}{2})\bar{\Lambda}(w_k)\bar{\Lambda}(w_k-\eta),\quad k=1,\cdots,N+2,\label{Rationa-BAE-2}
\eea
where $\Delta(u)$ is the quantum determinant with homogeneous limit $\{\theta_j=0\}$
\bea
\Delta(u)=(u-\eta)(u+\eta)(u-p)(u+p)(\sqrt{1+\xi^2}\,u+q)(\sqrt{1+\xi^2}\,u-q)(u+\eta)^{2N}(u-\eta)^{2N}.
\eea
The eigenvalue of the Hamiltonian (\ref{OBC}) ai determined by the solutions of above BAEs
\bea
E^o=\sum_{j=1}^{N+1}\frac{\eta^2}{\frac{\eta^2}{4}-z_j^2}-N.\label{XXX-Eigenvalue}
\eea

The hermitian of Hamiltonian (\ref{OBC}) requires that the boundary parameters satisfy
\bea
p^*=-p,\quad q^*=-q,\quad \xi^*=\xi. \label{XXX-Paramters-Hermitian-1}
\eea
The above constrains implies
\bea
R^{\dagger}(u)=-R(-u^*),\quad \lt(K^{(\pm)}(u)\rt)^{\dagger}=-K^{(\pm)}(-u^*),
\eea
which gives rise to the hermitian properties of the transfer matrix and its eigenvalue
\bea
\lt(t^o(u)\rt)^{\dagger}=t^o(-u^*),\quad \bar{\Lambda}^*(u)=\bar{\Lambda}(-u^*).\label{xxx-hermitian}
\eea

Combining the expansions (\ref{Ration-Expandions}) and (\ref{Ration-Expandions-1}), $t-W$ relation (\ref{Ration-Main-relation-1}) and the hermitian relation (\ref{xxx-hermitian}),
we conclude that if $z_j$ is a root of $\bar{\Lambda}(u)$, then $z^*_j$ must be the root and that
if $w_j$ is a root of $\bar{W}(u)$, then $w_j^*$ must be the root.

\subsection{Eigenfunctions in the thermodynamic limit at the ground state}
\begin{center}
\includegraphics{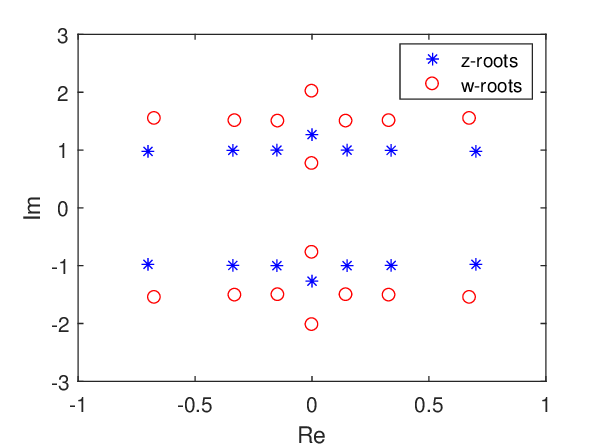}\\[5pt]
\parbox[c]{15.0cm}{\footnotesize{\bf Fig.~2.} The patterns of $z$-roots and $w$-roots in complex plane at the ground state with $N=6, \eta=i, p=-1.2i,\bar{q}=0.8i, \xi=1$.
The data are obtained by the exact numerical diagonalization. The blue asterisks indicate ${z}$-roots $\{z_j+\frac\eta2\}$ and red circles represent the ${w}$-roots $\{w_j\}$ with the inhomogeneous parameters $\{{\theta}_j=0\}$.}
\end{center}

As shown in figure 2, at the ground state, the $z$-roots and $w$-roots form the 2-strings. Besides the bulk conjugate pairs around the $\pm \eta$ lines, there exist boundary conjugate pairs at the imaginary axis. We can conclude that the $z$-roots form conjugate pairs as $\{\pm z_1 \eta, u_j^{(2)}\pm \eta|j=1,\cdots,N\}$ with real $z_1, u_j^{(2)}$, and $w$-roots form conjugate pairs as $\{\pm \chi_1 \eta, \pm \chi_2 \eta, w_j^{(2)}\pm \frac{3\eta}2|j=1,\cdots,N\}$ with real $\chi_1, \chi_2, w_j^{(2)}$. Substituting these 2-strings into Eqs.\eqref{Ration-Expandions} and \eqref{Ration-Expandions-1}, we obtain
\bea
\bar{\Lambda}_g(u)&=& 2\big(u-(z_1-\frac{1}{2})\eta\big)\big(u+(z_1+\frac{1}{2})\eta\big)
\prod_{j=1}^{N/2}(u-u_j^{(2)}-\frac{\eta}{2})(u+u_j^{(2)}-\frac{\eta}{2}) \no \\ && \times (u-u_j^{(2)}+\frac{3\eta}{2})(u+u_j^{(2)}+\frac{3\eta}{2}) \no \\
&=&2\big(u-(z_1-\frac{1}{2})\eta\big)\big(u+(z_1+\frac{1}{2})\eta\big) e^{2N( \bar{\lambda}_g^{(0)}(u)+\frac1{2N} \bar{\lambda}_g^{(1)}(u)+O(\frac1{N^2}) )}, \label{t-gs-o}\\[4pt]
\bar{W}_g(u)&=&(\xi^2-3)(u-\chi_1\eta)(u+\chi_1\eta) (u-\chi_2\eta) (u+\chi_2\eta)
\prod_{j=1}^{N/2} (u-w_j^{(2)}-\frac32 \eta)(u+w_j^{(2)}-\frac32 \eta) \no \\ && \times(u-w_j^{(2)}+\frac32 \eta)(u+w_j^{(2)}+\frac32 \eta) \no \\
&=&(\xi^2-3)(u-\chi_1\eta)(u+\chi_1\eta) (u-\chi_2\eta) (u+\chi_2\eta)e^{2N( \bar{\omega}_g^{(0)}(u)+\frac1{2N} \bar{\omega}_g^{(1)}(u)+O(\frac1{N^2}) )},\label{w-gs-o} \\
\bar{\Lambda}_g(0)&=&2(z_1-\frac12)(z_1+\frac12)e^{2N( \bar{\lambda}_g^{(0)}(u)+\frac1{2N} \bar{\lambda}_g^{(1)}(u)+O(\frac1{N^2}) )} =2\,p\,q\equiv2\,p\,\bar{q}\sqrt{1+\xi^2},
\eea
where the leading terms are determined by
\bea
&&\frac{\partial}{\partial u}\bar{\lambda}_g^{(\beta)}(u)=\int_{-\infty}^{\infty} (\frac{1}{u-\lambda-\frac\eta2}+\frac{1}{u-\lambda+\frac{3\eta}2}) \rho^{(\beta)}_g(\lambda)\mathrm{d}\lambda,\quad \beta=0,1,\label{lam-o-d}\\
&&\frac{\partial}{\partial u}\bar{\omega}_g^{(\beta)}(u)=\int_{-\infty}^{\infty} \bigg(\frac{1}{u-\lambda-\eta}+\frac{1}{u-\lambda+\eta} +\frac{1}{u-\lambda-2\eta}+\frac{1}{u-\lambda+2\eta}\bigg) \rho^{(\beta)}_w(\lambda)\mathrm{d}\lambda,\,\beta=0,1.\label{w-gs-d}
\eea

Taking the difference of \eqref{t-theta-o} at two inhomogeneous points and considering the continuum limit, we have
\bea
\frac{\partial}{\partial u}\ln\big[(u+\frac\eta2)(u-\frac\eta2)\bar{\Lambda}(u)\bar{\Lambda}(u-\eta)\big] =\frac{\partial}{\partial u} \ln\Delta(u)+O(\frac1N).\label{lamlamo}
\eea
Similarly with the periodic boundary case, substituting \eqref{t-gs-o} into \eqref{lamlamo}, we have
\bea\label{lamlam2-o}
&&\frac{\partial}{\partial u}[\bar{\lambda}_g^{(0)}(u)+\bar{\lambda}_g^{(0)} (u-\eta)]=\frac{1}{u+\eta}+\frac{1}{u-\eta},\qquad \bar{\lambda}_g^{(0)} (0)=0, \no \\
&&\frac{\partial}{\partial u}[\bar{\lambda}_g^{(1)}(u)+\bar{\lambda}_g^{(1)} (u-\eta)]=\frac1{u+\eta}+\frac1{u-\eta}+\frac1{u+p\eta}+\frac1{u-p\eta} +\frac1{u+\bar{q}\eta}+\frac1{u-\bar{q}\eta} -\frac1{u+\frac\eta2}- \frac1{u-\frac\eta2} \no \\ && \qquad -\frac1{u+\frac\eta2}-\frac1{u-\frac\eta2} -\frac1{u+(z_1-\frac12)\eta}-\frac1{u-(z_1-\frac12)\eta} -\frac1{u+(z_1+\frac12)\eta}-\frac1{u-(z_1+\frac12)\eta}, \no \\ &&
\bar{\lambda}_g^{(1)} (0)=\ln(2\,p\,\bar{q}\sqrt{1+\xi^2}) -\ln\big[2(z_1+\frac12)(z_1-\frac12)\big].
\eea
Substituting \eqref{lam-o-d} into \eqref{lamlam2-o} and solving it by the Fourier transformation, we obtain the densities of $z$-roots at the ground state as
\bea
&&\tilde{\rho}^{(0)}_g(w)=\frac{a_2(w)}{2\pi\big[a_1(w)+a_3(w)\big]}, \no \\
&&\tilde{\rho}^{(1)}_g(w)=\frac{1}{2\pi\big[a_1(w)+a_3(w)\big]}\big[a_2(w)+a_{2p}(w) +a_{2\bar{q}}(w)-a_1(w)-a_{2z_1-1}(w)-a_{2z_1+1}(w)\big],\label{density-o}
\eea
where $\tilde{\rho}^{(\beta)}_g(w)$ is the Fourier transformation of $\rho^{(\beta)}_g(\lambda)$ and $a_n(w)=e^{-n|w|/2}$.
By incorporating the 2-string structure into the energy expression \eqref{XXX-Eigenvalue} and integrating it with the densities \eqref{density-o} as the weighting factor, we obtain the ground state energy
\bea
E_g^o&=&-N i \int_{-\infty}^{\infty}\bigg(\frac1{z+\frac i2}+\frac1{z-\frac {3i}2} \bigg) \int^{\infty}_{-\infty}e^{i\omega z}\big(\tilde{\rho}^{(0)}_g(\omega)+\tilde{\rho}^{(1)}_g(\omega)\big)\, \mathrm{d} \omega \, \mathrm{d} z +\frac1{z_1^2-\frac 14} - N\no \\ [4pt]
&=&(1-4\ln2)N-1+\pi-2\ln2+\frac 1{|p|}+\frac {\sqrt{1+\xi^2}}{|q|}-2\,i \int_0^\infty\frac{e^{-|p|w}+e^{-|q|w/\sqrt{1+\xi^2}}}{1+e^{-w}}.
\eea
Substituting the densities \eqref{density-o} into \eqref{lam-o-d}, we arrive at
\bea\label{lam0-o}
\bar{\lambda}_g^{(0)}(u)&=&\ln\frac{2\Gamma(1+\frac{iu}2) \Gamma(\frac32-\frac{iu}2)}{\Gamma(\frac12+\frac{iu}2) \Gamma(1-\frac{iu}2)}, \\
\bar{\lambda}_g^{(1)}(u)&=&\ln\left[\frac{4\sqrt{1+\xi^2}}{\big(u-(z_1 -\frac12)\eta\big) \big(u+(z_1+\frac12)\eta\big)(u+\frac\eta2)} \frac{\cosh(\frac{\pi u}2-\frac{i\pi}4)} {\sinh(\frac{\pi u}2-\frac{i\pi}4)} \frac{\Gamma(1+\frac{iu}2) \Gamma(\frac32-\frac{iu}2)} {\Gamma(\frac12+\frac{iu}2) \Gamma(1-\frac{iu}2)}\right. \no \\ && \left.\times
\frac{\Gamma(\frac{p+1}2+\frac{iu}2) \Gamma(\frac{p+2}2-\frac{iu}2)} {\Gamma(\frac{p}2+\frac{iu}2) \Gamma(\frac{p+1}2-\frac{iu}2)}
\frac{\Gamma(\frac{\bar{q}+1}2+\frac{iu}2) \Gamma(\frac{\bar{q}+2}2-\frac{iu}2)} {\Gamma(\frac{\bar{q}}2+\frac{iu}2) \Gamma(\frac{\bar{q}+1}2-\frac{iu}2)}\right],
\eea
which implies
\bea
\bar{\Lambda}_g(u)&=&\frac{8\sqrt{1+\xi^2}}{u+\frac\eta2} \frac{\cosh(\frac{\pi u}2-\frac{i\pi}4)} {\sinh(\frac{\pi u}2-\frac{i\pi}4)} \frac{\Gamma(1+\frac{iu}2) \Gamma(\frac32-\frac{iu}2)} {\Gamma(\frac12+\frac{iu}2) \Gamma(1-\frac{iu}2)}
\frac{\Gamma(\frac{p+1}2+\frac{iu}2) \Gamma(\frac{p+2}2-\frac{iu}2)} {\Gamma(\frac{p}2+\frac{iu}2) \Gamma(\frac{p+1}2-\frac{iu}2)}
\no \\ && \times \frac{\Gamma(\frac{\bar{q}+1}2+\frac{iu}2) \Gamma(\frac{\bar{q}+2}2-\frac{iu}2)} {\Gamma(\frac{\bar{q}}2+\frac{iu}2) \Gamma(\frac{\bar{q}+1}2-\frac{iu}2)} \bigg(\frac{2\Gamma(1+\frac{iu}2) \Gamma(\frac32-\frac{iu}2)}{\Gamma(\frac12+\frac{iu}2) \Gamma(1-\frac{iu}2)}\bigg)^{2N}e^{O(\frac1N)}.
\eea

Now, we consider the leading terms of the $\bar{W}_g(u)$ in the thermodynamic limit.
Analogous to the case of periodic boundary condition, we set the variable $z_j=u_j^{(2)}+\eta$ in BAEs \eqref{Rationa-BAE-1} and multiply it by its conjugate counterpart. As a result, we obtain
\bea\label{delta-w-o}
\Delta(u_j^{(2)}+\frac{\eta}{2})\Delta(u_j^{(2)}-\frac{\eta}{2}) =(u_j^{(2)}+\frac{\eta}{2})^{2N+2}(u_j^{(2)}-\frac{\eta}{2})^{2N+2}\, \bar{W}(u_j^{(2)}+\frac{\eta}{2})\bar{W}(u_j^{(2)}-\frac{\eta}{2}).
\eea
By taking the derivative of the logarithm of \eqref{delta-w-o}, we obtain
\bea\label{ww1-o}
&&\frac1{2N} \frac{\partial}{\partial u} \ln [W(u+\frac\eta2) W(u-\frac\eta2)]=\frac1{2N}\frac{\partial}{\partial u} \bigg(\ln \big[\Delta(u+\frac{\eta}{2})\Delta(u-\frac{\eta}{2})\big] \no \\ && \qquad -\ln\big[(u+\frac{\eta}{2})^{2N+2}(u-\frac{\eta}{2})^{2N+2}\big]\bigg) +O(\frac1{N^2}).
\eea
Substituting \eqref{w-gs-o} into \eqref{ww1-o}, we have
\bea\label{ww2-o}
&&\hspace{-0.5truecm}\frac{\partial}{\partial u}\big(\bar{\omega}_g^{(0)}(u+\frac\eta2)+\bar{\omega}_g^{(0)} (u-\frac\eta2)\big)=\frac{1}{u+\frac{3\eta}{2}}+\frac{1}{u-\frac{3\eta}{2}}, \no \\
&&\hspace{-0.5truecm}\frac{\partial}{\partial u}\big(\bar{\omega}_g^{(1)}(u+\frac\eta2)+\bar{\omega}_g^{(1)} (u-\frac\eta2)\big)=\frac1{u+\frac32\eta}+\frac1{u-\frac32\eta} +\frac1{u+(p+\frac12)\eta}+\frac1{u-(p+\frac12)\eta}
\no \\ &&
+\frac1{u+(p-\frac12)\eta}+\frac1{u-(p-\frac12)\eta} +\frac1{u+(\bar{q}+\frac12)\eta}+\frac1{u-(\bar{q}+\frac12)\eta} +\frac1{u+(\bar{q}-\frac12)\eta}+\frac1{u-(\bar{q}-\frac12)\eta}
\no \\ &&
-\bigg( \frac1{u+\frac12\eta}+\frac1{u-\frac12\eta} +\frac1{u+(\chi_1+\frac12)\eta}+\frac1{u-(\chi_1+\frac12)\eta} +\frac1{u+(\chi_1-\frac12)\eta}+\frac1{u-(\chi_1-\frac12)\eta}
\no \\ &&
+\frac1{u+(\chi_2+\frac12)\eta}+\frac1{u-(\chi_2+\frac12)\eta} +\frac1{u+(\chi_2-\frac12)\eta}+\frac1{u-(\chi_2-\frac12)\eta}
\bigg).
\eea
Substituting \eqref{w-gs-d} into \eqref{ww2-o} and solving it by the Fourier transformation, we obtain the densities of $w$-roots at the ground state
\bea
&&\tilde{\rho}^{(0)}_w(w)=\frac{a_2(w)}{2\pi\big[a_1(w)+a_3(w)\big]},\label{w-o-d} \\
&&\tilde{\rho}^{(1)}_w(w)=\frac1{2\pi\big[a_1(w)+a_3(w)\big]}\big[a_3(w)+a_{2p+1}(w) +a_{2p-1}(w)+a_{2\bar{q}+1}(w)+a_{2\bar{q}-1}(w) \no \\
&& \qquad -a_{1}(w)-a_{2\chi_1-1}(w) -a_{2\chi_1+1}(w)-a_{2\chi_2-1}(w) -a_{2\chi_2+1}(w) \big].\label{w-o-d1}
\eea
Substituting the densities \eqref{w-o-d} and \eqref{w-o-d1} into \eqref{w-gs-d}, we arrive at
\bea\label{w0-o}
\frac{\partial}{\partial u} \bar{\omega}_g^{(0)}(u) &=&\frac{\partial}{\partial u} \ln\frac{\Gamma(\frac32+\frac{iu}2) \Gamma(\frac32-\frac{iu}2)}{\Gamma(1+\frac{iu}2) \Gamma(1-\frac{iu}2)}
=\frac{\partial}{\partial u} \ln \bigg[\frac{(u+\eta)(u-\eta)}{2u} \tanh\frac{\pi u}2\bigg], \\
\frac{\partial}{\partial u} \bar{\omega}_g^{(1)}(u) &=& \frac{\partial}{\partial u} \ln \bigg( \frac{(u-p\eta)(u+p\eta)(u-\bar{q}\eta) (u+\bar{q}\eta)}{(u-\chi_1\eta)(u+\chi_1\eta)(u-\chi_2\eta)(u+\chi_2\eta)}
\frac{\Gamma(\frac{3}2+\frac{iu}2) \Gamma(\frac{3}2-\frac{iu}2)} {\Gamma(1+\frac{iu}2) \Gamma(1-\frac{iu}2)} \no \\ && \times
\frac{\Gamma(\frac{1}2+\frac{iu}2) \Gamma(\frac{1}2-\frac{iu}2)} {\Gamma(1+\frac{iu}2) \Gamma(1-\frac{iu}2)}\bigg).
\eea
Then the leading terms can be obtained by taking the integration of above equation and the finial results are
\bea\label{w01-o}
\bar{\omega}_g^{(0)}(u) &=& \ln\bigg( \frac{\bar{C}_w^{(0)}}{2}\frac{(u+\eta)(u-\eta)}{u} \tanh\frac{\pi u}2\bigg), \\
\bar{\omega}_g^{(1)}(u) &=& \ln\bigg( \frac{(u-p\eta)(u+p\eta)(u-\bar{q}\eta) (u+\bar{q}\eta)}{(u-\chi_1\eta)(u+\chi_1\eta)(u-\chi_2\eta)(u+\chi_2\eta)}
\frac{\Gamma(\frac{3}2+\frac{iu}2) \Gamma(\frac{3}2-\frac{iu}2)} {\Gamma(1+\frac{iu}2) \Gamma(1-\frac{iu}2)}
\frac{\Gamma(\frac{1}2+\frac{iu}2) \Gamma(\frac{1}2-\frac{iu}2)} {\Gamma(1+\frac{iu}2) \Gamma(1-\frac{iu}2)}\bigg) \no \\ && +\ln \bar{C}_w^{(1)}.
\eea
Thus the the ground state eigenfunction $\bar{W}_g(u)$ in the thermodynamic limit can be expressed as
\bea\label{wg-o}
\bar{W}_g(u) &=& 4\bar{C}_w^{(1)}(\xi^2-3)(u-p\eta)(u+p\eta)(u-\bar{q}\eta)(u+\bar{q}\eta) \tanh^2\frac{\pi u}2
\no \\ && \times \frac{(u+\eta)^{2N+1}(u-\eta)^{2N+1}}{u^{2N+2}} \bigg(\frac{\bar{C}_w^{(0)}}{2}\tanh\frac{\pi u}2\bigg)^{2N}.
\eea
Substituting Eq.\eqref{wg-o} into the $t-W$ relation \eqref{tw-o}, we have
\bea\label{tw1-o}
&&(u+\frac{\eta}{2})\,(u-\frac{\eta}{2})\bar{\Lambda}_g(u)\bar{\Lambda}_g(u-\eta) = (1+\xi^2)(u-p\eta)(u+p\eta)(u-\bar{q}\eta)(u+\bar{q}\eta) (u-\eta)^{2N+1}(u+\eta)^{2N+1} \no \\ && \quad \times \bigg\{1-\frac{4(\xi^2-3)}{1+\xi^2} \bar{C}_w^{(1)} \tanh^2\frac{\pi u}2 \bigg( \frac{\bar{C}_w^{(0)}}{2}\tanh\frac{\pi u}2\bigg)^{2N} e^{O(\frac1N)}\bigg\}.
\eea
When $u$ tends to infinity, the coefficient of the highest order term of the left hand side of \eqref{tw1-o} is 4. Thus the corresponding coefficient of the right
hand side of \eqref{tw1-o} should be also 4, which gives $\bar{C}_w^{(0)}=2$ and $\bar{C}_w^{(1)}=\frac14$.
Because $\tanh\frac{\pi u}2<1$, the second term of the right hand side of \eqref{tw1-o} is negligible in the thermodynamic limit.

\section{Conclusion}

In this paper, we take the XXX spin chain as an example to study the $t-W$ scheme for the quantum integrable systems.
We present the exact solutions of the model with periodic and generic open boundary conditions.
We also obtain the analytical expressions of the ground state eigenfunctions of the transfer matrix and $\mathbb{W}$ operator in the thermodynamic limit.
By analyzing these expressions, we find that the ratio of the quantum determinant with the $W$ function converges to zero when the number of size tends to infinity. Thus the main contribution in the
$t-W$ relation comes from the quantum determinant. This finding serves as a compelling proof of the validity of the extensively applied inversion relation
in the field of integrability.

\addcontentsline{toc}{chapter}{Appendix A: Appendix section heading}

\section*{Appendix A: Proof of the $t-W$ relation  }
\setcounter{equation}{0}
\renewcommand{\theequation}{A.\arabic{equation}}

Starting from the YBE (\ref{QYB}) with certain shift of spectral parameter and
using the fusion technique \cite{kul81, kir86}, we obtain
\bea
R_{2,3}(u)R_{1,3}(u-\eta)P^{(-)}_{1,2}
=P^{(-)}_{1,2}R_{2,3}(u)R_{1,3}(u-\eta)P^{(-)}_{1,2}
=(u+\eta)(u-\eta)\times {\rm id}.\label{Q-determinant}
\eea
We see that the fusion result of $R$-matrices with one-dimensional antisymmetric projector $P^{(-)}_{1,2}$ is a number.
The fusion of monodromy matrices with $P^{(-)}_{1,2}$ gives
\bea
P^{(-)}_{1,2}T_2(u)\,T_1(u-\eta)P^{(-)}_{1,2}=a(u)\,d(u-\eta)\times {\rm id}.
\eea
Then we obtain the first term in the very relation (\ref{t-W-relation-op}).

The fusion of $R$-matrices with the three-dimensional symmetric projector $P^{(+)}_{1,2}$ gives
\bea
P^{(+)}_{1,2}\,R_{2,3}(u)\,R_{1,3}(u-\eta)\,P^{(+)}_{1,2}
=u\times  R^{(1,\frac{1}{2})}_{\{1,2\},3}(u),\label{Fusion-R}
\eea
where $R^{(1,\frac{1}{2})}_{\{1,2\},3}(u)$ is the $6\times 6$ fused $R$-matrix with the form of
\bea
R^{(1,\frac{1}{2})}_{\{1,2\},3}(u)=\lt(\begin{array}{cccccc}
u+\eta&&&&&\\[6pt]
&u-\eta&\sqrt{2}\eta&&&\\[6pt]
&\sqrt{2}\eta&u&&&\\[6pt]
&&&u&\sqrt{2}\eta&\\[6pt]
&&&\sqrt{2}\eta&u-\eta&\\[6pt]
&&&&&u+\eta
\end{array}\rt).\label{Fusion-R-1}
\eea
Then the fusion of monodromy matrices reads
\bea
P^{(+)}_{1,2}T_2(u)\,T_1(u-\eta)P^{(+)}_{1,2}=d(u)T^{(1,\frac{1}{2})}_{\{1,2\}}(u),\label{Fused-Mono}
\eea
where $T^{(1,\frac{1}{2})}_{\{1,2\}}(u)$ is the fused monodromy matrix which is constructed by the fused $R$-matrix $R^{(1,\frac{1}{2})}_{\{1,2\},j}(u)$ as
\bea
T^{(1,\frac{1}{2})}_{\{1,2\}}(u)=R^{(1,\frac{1}{2})}_{\{1,2\},N}(u-\theta_N)\cdots R^{(1,\frac{1}{2})}_{\{1,2\},1}(u-\theta_1).\label{Fused-Mono-1}
\eea
Then we arrive at that the $\mathbb{W}(u)$ operator in Eq.(\ref{t-W-relation-op}) is
the fused transfer matrix as
\bea
\mathbb{W}(u)=tr_{\{1,2\}}T^{(1,\frac{1}{2})}_{\{1,2\}}(u).
\eea
From the constructions (\ref{Fusion-R-1}) and (\ref{Fused-Mono-1}), we know that the $\mathbb{W}(u)$ operator is a operator polynomial
of $u$ with the degrees $N$.

\section*{Appendix B: Hermitian property of the transfer matrix }
\setcounter{equation}{0}
\renewcommand{\theequation}{B.\arabic{equation}}

The $R$-matrix (\ref{R-matrix}) has the hermitian relation
\bea
R^{\dagger}_{0,j}(u)=-R_{0,j}(-u^*). \label{R-Hermitian}
\eea
Based on it, we obtain the hermitian conjugation of the transfer matrix in the homogeneous limit as
\bea
t^{\dagger}(u)&=&tr_0\lt\{\lt[R_{0,N}(u)\cdots R_{0,1}(u)\rt]^{\dagger}\rt\} \no\\[4pt]
&=&tr_0\lt\{R^{\dagger}_{0,1}(u)\cdots R^{\dagger}_{0,N}(u)\rt\} \no\\[4pt]
&\stackrel{(\ref{R-Hermitian})}{=}&(-1)^{N}tr_0\lt\{R_{0,1}(-u^*)\cdots R_{0,N}(-u^*)\rt\} \no\\[4pt]
&=&(-1)^{N}tr_0\lt\{R^{t_0}_{0,N}(-u^*)\cdots R^{t_0}_{0,1}(-u^*)\rt\} \no\\[4pt]
&\stackrel{(\ref{XXX-R-property})}{=}&tr_0\lt\{\sigma^y_0R_{0,N}(u^*-\eta)\cdots R_{0,1}(u^*-\eta)\,\sigma^y_0\rt\} \no\\[4pt]
&=&tr_0\lt\{R_{0,N}(u^*-\eta)\cdots R_{0,1}(u^*-\eta)\rt\}=t(u^*-\eta). \label{transfer-hermitian-1}
\eea
With the help of the $t-W$ relation (\ref{t-W-relation-op}), we find that the $\mathbb{W}(u)$ operator satisfies
\bea
\mathbb{W}^{\dagger}(u)=\mathbb{W}(u^*). \label{transfer-hermitian-2}
\eea
The relations (\ref{transfer-hermitian-1}) and (\ref{transfer-hermitian-2}) imply that the corresponding eigenvalues have the properties
\bea
\Lambda^*(u)=\Lambda(u^*-\eta),\quad W^*(u)=W(u^*).\label{T-W-eig-Hermitian-1}
\eea
Combining the relations (\ref{T-W-eig-Hermitian-1}) and (\ref{t-W-relation-Eigen})-(\ref{Expansion-4}), we conclude that if $z_j$
is the solution of $\Lambda(u)$, its complex conjugation $z^*_j$ must be the solution, and if
$w_j$ is a solution of $W(u)$, the $w^*_j$ must be the solution.

\end{document}